\documentclass[12pt,letterpaper,a4paper]{article}

\usepackage[includeheadfoot, 
            marginratio={1:1,2:3}, 
            width=412pt, 
            height=688pt,]{geometry}

\usepackage{amsmath}
\usepackage{amsfonts}
\usepackage{amssymb}
\usepackage{amsthm,amscd}

\usepackage{empheq}
\usepackage[margin=10pt,font=small]{caption}
\usepackage{stmaryrd}
\usepackage{color}
\usepackage{dsfont}
\usepackage{simplewick}
\usepackage{multirow}

\newcommand{\nc}{\newcommand}
\nc{\beq}{\begin{equation}}
\nc{\eeq}{\end{equation}}
\nc{\bea}{\begin{eqnarray}}
\nc{\eea}{\end{eqnarray}}
\def\ov{\overline}

\nc{\CORR}[1]{\big\langle \! #1 \! \big\rangle}
\nc{\NO}[1]{:\!\! #1 \!\! :}
\nc{\lb}{\llbracket}
\nc{\rb}{\rrbracket}
\nc{\bnabla}{\blacktriangledown}

\newcommand{\eq}[1]{\begin{equation}
                     \begin{split} #1 \end{split}
                     \end{equation}}

\newcommand{\corr}[1]{\bigl\langle #1 \bigr\rangle}
\newcommand{\noo}[1]{:\! #1 \!:\,}

\theoremstyle{definition}

\begin{document}

\vspace*{-1.5cm}

\begin{flushright}
  {\small
  MPP-2014-16\\
  LMU-ASC 03/14
  }
\end{flushright}

\vspace{1.5cm}

\begin{center}
  \textbf{\LARGE
A Note on the CFT Origin of the \\[0.2cm]
Strong Constraint of DFT
  }
\end{center}

\vspace{0.4cm}

\begin{center}
  Andr\'e Betz$^1$, Ralph Blumenhagen$^{1}$, Dieter L\"ust$^{1,2}$ and Felix Rennecke$^{1}$
\end{center}


\begin{center} 
\emph{$^{1}$ Max-Planck-Institut f\"ur Physik (Werner-Heisenberg-Institut), \\ 
   F\"ohringer Ring 6,  80805 M\"unchen, Germany } \\[0.1cm] 
\vspace{0.25cm}
\emph{$^{2}$ Arnold Sommerfeld Center for Theoretical Physics, LMU \\
	Theresienstra\ss e 37, 80333 M\"unchen, Germany}  \\[0.1cm] 
\end{center} 
\vspace{0.4cm}

\begin{abstract}
In double field theory, motivated by its field theoretic consistency, the
level matching condition is generalized to the so-called strong constraint. In
this note, it is investigated what the two-dimensional conformal field theory
origin of this constraint is. Initially treating the left- and right-movers as
independent, we compute the 
torus partition function as well as a generalized Virasoro-Shapiro amplitude. 
In non-compact directions the strong constraint arises from the
factorization of the Virasoro-Shapiro amplitude over physical
states as determined by the modular invariant partition function.
From the same argument,  along internal toroidal directions, 
no analogous  constraint arises.
\end{abstract}


\clearpage


\section{Introduction}
The T-duality symmetries, present in toroidal string compactifications, motivated various attempts to construct duality invariant theories. In 1990 Tseytlin realized T-duality as a world-sheet symmetry by treating left- and right-moving degrees of freedom on equal footing \cite{Tseytlin:1990nb,Tseytlin:1990va}: The bosonic field describing a  two-dimensional string world-sheet embedded into the target space-time  splits into a left and a right moving component. Then, T-duality acts as a reflection of the right-moving component \cite{Rocek:1991ps}, giving rise
to a winding coordinate.  

In a geometric target space approach to the problem pursued in
\cite{Hull:2004in,Dabholkar:2005ve,Hull:2006qs,Hull:2006va}, the usual and the
winding coordinates were considered as coordinates of a doubled manifold,
termed doubled geometry. Many quantum aspects of this theory were studied
further in \cite{Berman:2007vi,Berman:2007xn,Berman:2007yf}, where in
particular an $O(d,d)$ invariant target space effective action was
presented. Double field theory (DFT) was developed in \cite{Siegel:1993xq,Siegel:1993th,Hull:2009zb,Hohm:2010jy} as a covariant doubled target space approach to duality symmetries (recent reviews can be found in \cite{Aldazabal:2013sca,Berman:2013eva,Hohm:2013bwa}). Whereas in the doubled geometry approach the compact part of space is doubled, in double field theory the whole space-time manifold is doubled.

In order to reduce the doubled degrees of freedom to match with the physical
ones, common to all these approaches is the necessity of imposing
constraints. Since in DFT one treats the massless modes of the closed
string, the level matching condition $L_0-\overline L_0=0$ must be
satisfied. This leads to the so-called weak constraint
\eq{
          \partial_a \tilde\partial^a f=0
}
with $\partial_a$ and  $\tilde{\partial}^a$ derivatives with respect to the standard coordinates $x^a$ and the winding coordinates
$\widetilde{x}_a$, respectively. However, for 
consistency of DFT, i.e. in particular for the
closure of the symmetry algebra of generalized diffeomorphisms, 
a stronger version of this constraint is imposed  \cite{Siegel:1993th,Hohm:2010jy}, namely 
\eq{\label{dSC}
	\partial_a f\,\tilde{\partial}^a g+\tilde{\partial}^af\,\partial_a
        g=0\, 
}
for all functions $f,g$ depending on the doubled coordinates.
However, it turned out that this  \emph{ad hoc} introduced \emph{strong
  constraint} \eqref{dSC} is merely a sufficient condition for consistency.
In the so-called flux formulation of the DFT 
\cite{Hohm:2010xe,Hohm:2011ex,Aldazabal:2011nj,Geissbuhler:2011mx,Grana:2012rr, Geissbuhler:2013uka}, motivated
by the tetrad formalism of general relativity and the early work \cite{Siegel:1993xq,Siegel:1993th}, it was shown that
a weaker constraint, namely the so-called closure constraint, is also
sufficient for consistency of DFT.
This is supported by the observation that Scherk-Schwarz reductions
\cite{Scherk:1974ca} of DFT lead to consistent
gauged supergravity theories in lower dimensions \cite{Grana:2012rr} without 
implementing the strong constraint along the compact directions in an obvious
way (see \cite{Hassler:2014sba} for a recent discussion about compactification of DFT on non-geometric backgrounds). 

Following Tseytlin's approach \cite{Tseytlin:1990nb,Tseytlin:1990va}, in this
note a simple T-duality invariant CFT is used to study the string theoretic 
origin of the
strong constraint \eqref{dSC} for non-compact and toroidally compactified
spaces. We compute the 1-loop torus partition function as well as a duality invariant
version of the Virasoro-Shapiro amplitude for a general non-compact and a
Narain-compactified target space. For non-compact spaces, modular invariance
of the partition function leads to strong restrictions upon the space of
physical states. We show that, provided that the single poles  of the
generalized Virasoro-Shapiro correspond to the mass spectrum of the theory,
the strong constraint follows. Hence the strong constraint is a consequence of
the relative consistency of modular invariance of the vacuum one-loop 
diagram and the pole structure of tree-level string scattering amplitudes.

Performing the same analysis for a toroidal compactification, no analogous
restriction can be derived. As we do not consider fluxes, this can be considered a special case of the results in \cite{Grana:2012rr}, whose consistency conditions are restrictions put upon possible fluxes. Our findings are in favor of the conjecture that the strong constraint is necessary along non-compact directions, whereas consistency (such as closure of the gauge algebra) only requires weaker restrictions in the compact directions.

In order to provide  confidence that the duality invariant CFT we are using
is indeed related to DFT, in an appendix, we review the match of 
the CFT tree-level scattering amplitude of three massless states  
with the (effective) space-time action of DFT \cite{Hull:2009zb}. 

\section{T-duality invariant CFT}
Treating left- and right-movers on equal footing makes T-duality manifest. In the following the necessary CFT-setup will be introduced briefly. The relation of this theory to DFT is shown in the appendix. In particular, winding coordinates dual to the usual coordinates will be introduced despite the absence of compact directions, which will be discussed in section~\ref{sec:compact}. Therefore momentum and winding modes are not quantized.

\subsubsection*{The free boson and T-duality}
The free boson on an Euclideanized world-sheet $\Sigma$ with $\partial\Sigma=\emptyset$ is considered. Fixing conformal gauge allows the world-sheet metric to be $h=\mathrm{diag}(1,1)$ and the metric $G$ on the $d$-dimensional (non-compact) target space $M$ is assumed to be constant. 
In complex coordinates the sigma model reads
\eq{\label{Sfb}
	S=\frac{1}{2\pi\alpha'}\int_\Sigma \!dz\,d\bar{z}\,G_{ab}\,\partial X^a\,\bar{\partial}X^b \,.
}
The equations of motion of this action imply  conserved (anti-)ho\-lo\-mor\-phic currents $J^a(z) = i\partial X^a$ and $\bar J^a(\bar z) = i\bar\partial X^a$.
This allows for a splitting of the coordinates into a left- and right-moving part: $X^a(z,\bar z)=X_L^a(z)+X_R^a(\bar z)$. As for every CFT, the energy momentum tensor
also spits into a holomorphic  $T(z) = -\frac{1}{\alpha'} G_{ab}\partial X^a\partial X^b$ and an anti-holomorphic component $\bar T(\bar z) = -\frac{1}{\alpha'} G_{ab}\bar\partial X^a\bar\partial X^b$. The two-point function of the 
left and right moving coordinates are
\eq{\label{prop}
		\corr{X_L^a(z_1)X_L^b(z_2)} &= -\frac{\alpha'}{2}\,G^{ab}\ln z_{12}\\
		\corr{X_R^a(\bar z_1)X_R^b(\bar z_2)} &= -\frac{\alpha'}{2}\,G^{ab}\ln \bar z_{12}
}
with $z_{ij}=z_i-z_j$.

For a toroidal target space, T-duality is a reflection of the right-moving coordinates \cite{Rocek:1991ps}, i.e.
\eq{
	X^a(z,\bar z)=X_L^a(z)+ X_R^a(\bar z)
	\quad\xleftrightarrow{\mathrm{T-duality}}\quad
	\widetilde{X}^a(z,\bar z)=X_L^a(z) -X_R^a(\bar z) \,.
}
In particular the energy momentum tensor and the propagator \eqref{prop} are invariant under T-duality. Compactifying the free bosons on a torus, due to the new boundary conditions $X^a(e^{-2\pi i}z,e^{2\pi i}\bar z)=X^a(z,\bar z)+2\pi w^a$, in addition to momentum $p^a$  winding $w^a$ is introduced. 
These are related to the left- and right-moving momenta appearing in the mode expansion 
\eq{\label{modeexp}
	X_{L}^a(z) &= q_L^a -\frac{i\alpha'}{2}\,k_L^a\,\ln z
	+i\sqrt{\frac{\alpha'}{2}}\sum_{n\neq 0}\frac{\alpha_n^a}{nz^n} \\
	X_R^a(\bar z) &=q_R^a -\frac{i\alpha'}{2}\,k_R^a\,\ln\bar z
	+i\sqrt{\frac{\alpha'}{2}}\sum_{n\neq 0}\frac{\bar \alpha_n^a}{n\bar z^n}
}
as $p^a = \frac{1}{2}(k_L^a+k_R^a)$ and $w^a = \frac{1}{2}(k_L^a-k_R^a)$. 
For the zero modes usually describing the center of mass position of the string we also split these into  left and right moving components so that
$x^a=q^a_L+q^a_R$ and $\tilde x^a=q^a_L-q^a_R$.
Therefore T-duality exchanges momentum and winding. 

As proposed in  \cite{Tseytlin:1990nb,Tseytlin:1990va}, this  suggests
to consider  $X_L^a$ and $X_R^a$ as well as momentum and winding on equal footing generically, i.e. not only for a toroidal compactification. Note that the resulting theory is not governed by the sigma model \eqref{Sfb} anymore. To make T-duality manifest, it is convenient to introduce the propagators for standard
and winding coordinates
\eq{\label{props}
	\corr{X^a(z_1,\bar z_1)X^b(z_2,\bar z_2)} &= -\frac{\alpha'}{2}\,G^{ab}\ln|z_{12}|^2 \\
	\corr{\widetilde{X}^a(z_1,\bar z_1)\widetilde{X}^b(z_2,\bar z_2)} &= -\frac{\alpha'}{2}\,G^{ab}\ln|z_{12}|^2 \\
	\corr{X^a(z_1,\bar z_1)\widetilde{X}^b(z_2,\bar z_2)} 
		&= -\frac{\alpha'}{2}\,G^{ab}\ln\frac{z_{12}}{\bar z_{12}}\,.
}
In the following, we will determine elementary properties of this theory, at first, without referring to any compactified directions. The presence of compact directions will be studied in section~\ref{sec:compact}.

\subsubsection*{Vertex operators and descendants}
The manifest duality-invariant primary field solely containing the coordinate fields is
\eq{\label{dtvo}
	V_{p,w}(z,\bar z) &= \noo{e^{ip_aX^a(z,\bar z)}\,e^{iw_a\widetilde{X}^a(z,\bar z)}}\,,
}
which will be called tachyon in the following. It is a primary field of weight
\eq{\label{cw:V}
	(h,\bar h) = \left(\frac{\alpha '}{4}(p+w)^2,
		\frac{\alpha '}{4}(p-w)^2\right)\,.
}
The mass of such a state is
\eq{
             M^2=-{2\over \alpha'}(h+\bar h)=-(p^2+w^2)\, .
}
The OPE of two such fields is
\eq{\label{VVope}
	V_{p_1,w_1}(z_1,\bar z_1)\,V_{p_2,w_2}(z_2,\bar z_2)
	=&|z_{12}|^{\alpha'(p_1\cdot p_2+w_1\cdot w_2)}\,
		\Bigl(\frac{z_{12}}{\bar z_{12}}\Bigr)^{\frac{\alpha'}{2}(p_1\cdot w_2+w_1\cdot p_2)}
		\\&\times V_{p_1+p_2,w_1+w_2}(z_2,\bar z_2)+\dots\,,
}
and admits a logarithmic branch point whose absence (locality) requires the quantization condition	
\eq{
\label{localimply}
\alpha'(p_1\cdot w_2+w_1\cdot p_2)\in\mathbb{Z}\, .
} 
Let us also comment on the first descendant states of \eqref{dtvo}:
\begin{itemize}
\item At the first excited level one has a \emph{form field} $\mathcal{A}_{p,w}$ and its complex conjugate $\bar{\mathcal{A}}_{p,w}$
\eq{
	\mathcal{A}_{p,w}(z,\bar z) &= A_a\noo{\partial X^a(z)\,V_{p,w}(z,\bar z)}\\
	\bar{\mathcal{A}}_{p,w}(z,\bar z) &= \bar A_a\noo{\bar\partial X^a(\bar z)\,V_{p,w}(z,\bar z)}
}
with $A$ and $\bar A$ one-forms. For heterotic torus compactifications these states give rise to the well-known enhancement of the gauge group. $\mathcal{A}$ is primary with conformal weight $(h,\bar h)=(1+\frac{\alpha'}{4}(p+w)^2,\frac{\alpha'}{4}(p-w)^2)$ if it is transversely   polarized in the sense  $A_a(p^a+w^a) =0$. Similarly, $\bar{\mathcal{A}}$ is primary with $(h,\bar h)=(\frac{\alpha'}{4}(p+w)^2,1+\frac{\alpha'}{4}(p-w)^2)$ for $\bar A_a(p^a-w^a) =0$.
\item At the next level one finds a \emph{$(0,2)$-tensor field} $\mathcal{E}_{p,w}$
\eq{\label{gravVO}
	\mathcal{E}_{p,w}(z,\bar z) 
	= E_{ab}\noo{\partial X^a(z)\,\bar\partial X^b(\bar z)\,V_{p,w}(z,\bar z)}
}
with the polarization $E_{ab}$. It is a primary field with $(h,\bar h)=(1+\frac{\alpha'}{4}(p+w)^2,1+\frac{\alpha'}{4}(p-w)^2)$ for transverse  polarization in the sense $E_{ab}(p^a+w^a)=0=E_{ab}(p^b-w^b)$. 
\end{itemize}

In the appendix we show explicitly that string scattering amplitudes of three such states \eqref{gravVO} can be matched precisely with interactions in DFT. This gives credence to our  usage of this duality invariant CFT
as a two-dimensional world-sheet model of DFT.

The quantum version of the classical vanishing of the energy momentum tensor is the annihilation of a state by the generators of the conformal group (up to a normal-ordering constant for the generator of rescalings). As a consequence, \emph{physical states} have to be primary fields of conformal weight $(h,\bar h)=(1,1)$, i.e.  in particular they are level-matched. For the four states considered above the constraints are shown in table~\ref{tab:states}.
\begin{table}[h!]
\centering
\begin{tabular}{c||c|c|l}
 state& level-matching & primary & mass \\\hline
$V_{p,w}$ & $p\cdot w=0$ & --- & $M^2=-\frac{4}{\alpha'}$\\[0.1cm]
$\mathcal{A}_{p,w}$ & $p\cdot w =-\frac{1}{\alpha'}$ & $A_m(p^m+w^m)=0$ & $M^2=-\frac{2}{\alpha'}$ \\[0.1cm]
$\bar{\mathcal{A}}_{p,w}$ & $p\cdot w =\frac{1}{\alpha'}$ &$\bar A_m(p^m-w^m)=0$ & $M^2=-\frac{2}{\alpha'}$ \\[0.1cm]
$\mathcal{E}_{p,w}$ & $p\cdot w=0$ & $E_{mn}(p^m+w^m)=0= E_{mn}(p^n-w^n)$ & $M^2=0$
\end{tabular}
\caption{The physical state condition requires the operators to be level-matched primaries of conformal weight $(1,1)$. This sets the mass of the states.}
\label{tab:states}
\end{table}
\newline
\noindent
Clearly $V_{p,w}$ corresponds to a negative mass$^2$ state, i.e., as expected,  it is a  tachyon. Moreover, the two states 
$\mathcal{A}_{p,w}$ and $\bar{\mathcal{A}}_{p,w}$
are tachyonic as well.
Finally, $\mathcal{E}_{p,w}$ is massless and therefore,
depending on the polarization, gives the graviton, the $B$-field and the dilaton. 
Next we will consider the one-loop partition function whose modular invariance imposes additional  constraints relating the  holomorphic with the anti-holomorphic sector.  


\section{The one-loop partition function}

In this section, we will compute the torus partition function for the CFT introduced above and work out the modular properties in some detail.
For a CFT defined on the world sheet torus with modular parameter $\tau$ and Hilbert space $\mathcal{H}$, the partition function is given by
\eq{\label{Z}
	Z(\tau,\bar\tau)&=\mathrm{tr}_{\mathcal{H}}\bigl(q^{L_0-\frac{c}{24}}\,\bar
        q^{\bar L_0-\frac{c}{24}}\bigr)
}
with $q=e^{2\pi i\tau}$. The trace is taken over the whole Hilbert space
$\mathcal{H}$ which beyond the oscillator modes also 
includes the continuous trace over momenta and windings.
This  can be evaluated as
\eq{\label{mwint}
	f(\tau,\bar \tau)&=\langle p,w|p,w\rangle\,
 \frac{1}{2}\,
		\left(\int\frac{d^dk_L}{(2\pi)^d}\,e^{i\frac{\pi}{2}\alpha'\,k_L^2\,\tau}\right)
		\left(\int\frac{d^dk_R}{(2\pi)^d}\,e^{-i\frac{\pi}{2}\alpha'\,k_R^2\,\bar\tau}\right)\,.
}
The evaluation of the trace over the oscillator part is as usual so that
altogether we obtain
\eq{
	Z(\tau,\bar\tau) = \frac{f(\tau,\bar \tau)}{|\eta(\tau)|^{2d}}\, ,
}
which has to be modular invariant. For $\mathrm{Im}(\tau)>0$ the integral
\eqref{mwint} can be evaluated to be proportional to $|\tau|^{-d}$; this is
not invariant under a modular $T$-transformation $\tau\to\tau+1$ as $|\eta(\tau)|$ is invariant itself. $T$-invariance yields the level matching condition
\eq{\label{lmc}
	\alpha'\,p\cdot w\in\mathbb{Z}\quad\Longleftrightarrow\quad
	\frac{\alpha'}{4}(k_L^2-k_R^2)\in\mathbb{Z}\,,
}
i.e. the two integrals are not independent. Writing
$k_R^2=k_L^2-\frac{4}{\alpha'}m$ for an integer $m$, level matching can be
imposed by including a factor  $\delta(k_L^2-k_R^2-\frac{4}{\alpha'}m)$ in \eqref{mwint}. Then, to evaluate the remaining integral in \eqref{mwint} we introduce $d$-dimensional spherical coordinates with radius $|k_R|$; up to constant factors we are left with
\eq{\label{flm}
	f(\tau,\bar \tau) \sim
		e^{2\pi i\,m\,\tau}\int\frac{d^dk_L}{(2\pi)^d}\,|k_L|^{d-1}\,
		e^{-\pi\alpha'\,k_L^2\,\mathrm{Im}(\tau)}
	\sim \frac{\Gamma\left(d-\frac{1}{2}\right)}{\mathrm{Im}(\tau)^{\frac{d}{2}}}\,
		\frac{e^{2\pi i\,m\,
\tau}}{\mathrm{Im}(\tau)^{\frac{d-1}{2}}}
}
for $\mathrm{Im}(\tau)>0$, which is $T$-invariant. However, realizing that 
$\mathrm{Im}(\tau)^{\frac{d}{2}}|\eta(\tau)|^{2d}$ is already $S$-invariant, 
invariance under a modular $S$-transformation $\tau\to-\frac{1}{\tau}$ is
spoiled by the second factor     
$e^{2\pi i\,m\,
\tau}\,\mathrm{Im}(\tau)^{\frac{1-d}{2}}
$
in \eqref{flm}. 

For the unwanted factor in \eqref{flm} to be absent and for obtaining a modular invariant result, the second integral in \eqref{mwint} has to evaluate to
\eq{
	\int\frac{d^dk_R}{(2\pi)^d}\,e^{-i\frac{\pi}{2}\alpha'\,k_R^2\,\bar\tau}\,\delta(k_L,k_R)
	= g(\bar\tau)\,e^{-i\frac{\pi}{2}\alpha'\,k_L^2\,\bar\tau}\,.
}
$\delta(k_L,k_R)$ implements relations between the momenta and $g(\bar\tau)$ is a modular function independent of the momenta. Thus the modular function is given by
\eq{\label{mofu}
	g(\bar\tau)= \int\frac{d^dk_R}{(2\pi)^d}\,e^{i\frac{\pi}{2}\alpha'(k_L^2-k_R^2)\bar\tau}\,\delta(k_L,k_R)
	= e^{2\pi i m\,\bar\tau}\int\frac{d^dk_R}{(2\pi)^d}\,\delta(k_L,k_R)\,,
}
where we used level matching. In \eqref{mofu} $g(\bar\tau)$ factorizes into a
$\bar\tau$-dependent factor and a momentum dependent one. The former is not
modular invariant unless $m=0$. The remaining integral over the momentum must
be constant, i.e. $\delta(k_L,k_R)$ has to be of the form
$\delta^d(k_R-F(k_L))$, with $F$ a vector-valued function. 

We can determine $F$ as follows. Since $m=0$, level-matching \eqref{lmc} can be written as
\eq{
	\begin{pmatrix}k_L\\ k_R \end{pmatrix}^t
	\begin{pmatrix} \mathds{1} & 0 \\ 0 & -\mathds{1} \end{pmatrix}
	\begin{pmatrix}k_L\\ k_R \end{pmatrix} 
	\equiv \langle K,K\rangle_d=0
}
and is invariant under $O(d,d)$-transformations of the vector
$K=(k_L,k_R)^t$. Hence, to maintain level-matching while having a relation
between the left- and right-moving momentum requires them to be related by an
$O(d,d)$-transformation. Thus we can construct the general form of $F$ by
rotating the most simple solution $k_R=k_L$. 
An $O(d,d)$ transformation
$\mathcal{T}\in O(d,d)$ satisfies
$\mathcal{T}^t\mathrm{diag}(\mathds{1},-\mathds{1})\mathcal{T}=\mathrm{diag}(\mathds{1},-\mathds{1})$
so that in particular the transpose satisfies
\eq{\label{oddcond}
	\mathcal{T}^t=\begin{pmatrix}a^t&c^t\\b^t&d^t\end{pmatrix}\in O(d,d)
	\quad\Longleftrightarrow\quad
	\begin{cases}
	aa^t-bb^t=\mathds{1}\\
	cc^t-dd^t=-\mathds{1}\\
	ac^t-bd^t=0
	\end{cases}\,.
}
Acting with $\mathcal{T}$ on $(k_L,k_R)^t$ modifies the simple solution
according to
\eq{
	\{k_R=k_L\}\mapsto \{ck_L+dk_R =ak_L+bk_R\}
	\Longleftrightarrow \{k_R =(d-b)^{-1}(a-c)k_L\}\,.
}
Using the conditions \eqref{oddcond} for the matrix elements of
$\mathcal{T}^t$ we see that $(d-b)^{-1}(a-c)\in O(d)$. 
Therefore, the conditions for modular $T$- and $S$-invariance imply 
that the right and left momenta are related by an $O(d)$ transformation as
\eq{\label{micond}
	k_R= \mathcal{M}\,k_L \quad\mathrm{with}\quad\mathcal{M}\in O(d)\,.
}

Having showed that modular invariance requires the insertion of
$(2\pi)^d\delta^d(k_R-\mathcal{M}k_L)$ and denoting $\langle
p,w|p,w\rangle=V_d$, the final torus partition function reads
\eq{\label{Zmi}
	Z(\tau,\bar\tau)
	=\frac{V_d/2}{(2\pi\sqrt{\alpha'})^d\,\mathrm{Im}(\tau)^{d\over 2}\,|\eta(\tau)|^{2d}}\,.
}
Let us close this section with the following four remarks:
\begin{itemize}
\item In terms of momentum and winding, \eqref{micond} enforces $w=0$ for $\mathcal{M}=\mathds{1}$ and $p=0$ for $\mathcal{M}=-\mathds{1}$.
\item Invariance under a modular $T$-transformation implied 
   $\alpha' p\cdot w\in {\mathbb Z}$, while only the additional  
invariance under
   a modular $S$-transformation really led to   the weak constraint
  $p \cdot w=0$. 
\item The latter  truncates the spectrum as only those states are allowed
  whose number of left- and right- oscillator excitations match. Comparison
  with table~\ref{tab:states} therefore shows that in particular
  $\mathcal{A}_{p,w}$ and its complex conjugate are forbidden.
\item Locality implied 
$\alpha'(p_i\cdot w_j+w_i\cdot p_j)\in\mathbb{Z}$ but  the necessity  of the 
strong constraint $p_i\cdot w_j+w_i\cdot p_j=0$ does not follow from
modular invariance of the one-loop partition function. 
\end{itemize}

In string theory, the latter is related to the one-loop
vacuum polarization diagram with all string excitations running
in the loop. From our analysis so far it is clear that, 
in order to detect the origin 
of the strong constraint, one also  needs
to consider string diagrams  containing momenta and
winding of many states. For this reason, in the next section we consider
the string scattering amplitude of four tachyons.

\section{Tachyons scattering and the strong constraint}
\label{sec:scattering}

In the T-duality invariant CFT
the correlation function of $N$ tachyon vertex operators 
$V_{p_i,w_i}(z_i,\bar z_i)\equiv V_i$ can be straightforwardly computed as
\eq{\label{Ntc}
	\corr{V_1\dots V_N}=\!\!\!\!\!\prod_{1\leq i<j\leq N}
	|z_{ij}|^{\alpha'(p_i\cdot p_j+w_i\cdot w_j)}\,
	\Bigl(\frac{z_{ij}}{\bar z_{ij}}\Bigr)^{\frac{\alpha'}{2}(p_i\cdot
          w_j+w_i\cdot p_j)}\, \delta\Big({\textstyle \sum} p_i\Big)
        \delta\Big({\textstyle \sum} w_i\Big)\, .
}
The difference to the standard tachyon correlator is the $\frac{z_{ij}}{\bar
  z_{ij}}$-factor\,\footnote{$SL(2,\mathbb{C})$-invariance can be checked
  explicitly.}. In the following,  we will compute the duality invariant
Virasoro-Shapiro amplitude and study its pole structure.  

\subsubsection*{The duality invariant Virasoro-Shapiro amplitude}

The full string scattering amplitude of $N$ tachyons is given by
\eq{
\label{scatter}
   A_N(p_i,w_i)= g_s^N \, C_{S^2} \int \prod_{i=1}^N d^2 z_i \;
 &{\textstyle\prod_{j=1}^3 
    \delta(z_j-z_j^0)}\, |z_{12} z_{13} z_{23}|^2 \\
&\times 
             	\corr{V_1\dots V_N}(z_1,\ldots z_N)
}
Here the  conformal group $PSL(2,\mathbb{C})$ has been used to fix
three of the $N$ insertion points on the sphere.
The standard choice is  $z_1=0$, $z_2=1$ and $z_3\to\infty$.
Moreover, \eqref{scatter} includes the three $c$-ghost correlator
$\bigl|\corr{c(z_1)\,c(z_2)\,c(z_3)}\bigr|^2 = |z_{12}\,z_{23}\,z_{13}|^2$.
The prefactors are a factor of the closed string coupling constant   $g_c$ 
for every closed string vertex operator and $C_{S^2}$ accounting for various normalizations (see e.g. \cite{Blumenhagen:2013fgp}).

\vspace{0.3cm}
\noindent
{\it Three-point amplitude}
\medskip

\noindent
The three-tachyon amplitude is given by
\eq{
	A_3(p_i,w_i) = g_c^3\,C_{S^2}\,\corr{(c\,\bar c\, V_1)(c\,\bar c\, V_2)(c\,\bar c\, V_3)}
	=g_c^3\,C_{S^2}\,,
}
where the $\delta$-distributions implementing momentum and winding
conservation have to be understood as implicit. The three-point amplitude is
therefore identical to the standard  one for three tachyons without a winding dependence.

\vspace{0.3cm}
\noindent
{\it Four-point amplitude}
\medskip

\noindent
Using \eqref{Ntc} and reordering the monomials, the four-point amplitude reads
\eq{
	A_4(p_i,w_j) &= g_c^4\,C_{S^2}\int d^2z\,
	\corr{(c\,\bar c\, V_1)\,(c\,\bar c\, V_2)\,(c\,\bar c\, V_3)\,V_4}\\
	&=g_c^4\,C_{S^2}\int d^2z\Bigl\{
	z^{\alpha'(p_1\cdot w_4+w_1\cdot p_4)} (1-z)^{\alpha'(p_2\cdot w_4+w_2\cdot p_4)}\\
	&\hspace{2.8cm}\times
	|z|^{\alpha'(p_1-w_1)\cdot(p_4-w_4)}\,|1-z|^{\alpha'(p_2-w_2)\cdot(p_4-w_4)}\Bigr\}\,.
}
It is convenient to introduce two sets of Mandelstam variables
\eq{\label{mandel}
	s = -(k_{L3}+k_{L4})^2 \quad&, \quad\qquad \mathfrak{s}=-(k_{R3}+k_{R4})^2\\
	t = -(k_{L2}+k_{L4})^2 \quad&,\quad\qquad \mathfrak{t}=-(k_{R2}+k_{R4})^2\\
	u = -(k_{L1}+k_{L4})^2 \quad&,\quad\qquad \mathfrak{u}=-(k_{R2}+k_{R4})^2
}
with $s+t+u=\mathfrak{s}+\mathfrak{t}+\mathfrak{u}=-\frac{16}{\alpha'}$ by level matching and the mass-shell condition. The relation between the two sets is given by
\eq{\label{LRrel}
	(k_{Li}+k_{Lj})^2-(k_{Ri}+k_{Rj})^2 = 4(p_i\cdot w_j + w_i\cdot p_j)
	\in\frac{4}{\alpha'}\,\mathbb{Z}\,.
}
Defining the function $\alpha(s)=-1-\frac{\alpha'}{4}s$ the amplitude integrates to
\eq{
	A_4(p_i,w_j) = 2\pi\,g_c^4\,C_{S^2}\,
	\frac{\Gamma\bigl(\alpha(s)\bigr)\,\Gamma\bigl(\alpha(t)\bigr)\,\Gamma\bigl(\alpha(u)\bigr)}
	{\Gamma\bigl(\alpha(\mathfrak{t})+\alpha(\mathfrak{u})\bigr)\,
	\Gamma\bigl(\alpha(\mathfrak{s})+\alpha(\mathfrak{u})\bigr)
		\,\Gamma\bigl(\alpha(\mathfrak{s})+\alpha(\mathfrak{t})\bigr)}\,.
}
Using \eqref{LRrel}, the $\alpha$'s can be related as $\alpha(\mathfrak{s})=\alpha(s)-n_{34}$, where
\eq{\label{ncond}
	n_{ij}=\alpha'(p_i\cdot w_j + w_i\cdot p_j)\quad\mathrm{with}\quad n_{14}+n_{24} +n_{34} =0\,.
}
Then, in terms of the left-moving variables the amplitude becomes
\eq{\label{VS}
		A_4(p_i,w_j) = 
	\frac{2\pi\,g_c^4\,C_{S^2}\,\,
	\Gamma\bigl(\alpha(s)\bigr)\,\Gamma\bigl(\alpha(t)\bigr)\,\Gamma\bigl(\alpha(u)\bigr)}
	{\Gamma\bigl(\alpha(t)\!+\!\alpha(u)\!+\!n_{34}\bigr)\,
	\Gamma\bigl(\alpha(s)\!+\!\alpha(u)\!+\!n_{24}\bigr)\,
		\Gamma\bigl(\alpha(s)\!+\!\alpha(t)\!+\!n_{14}\bigr)}\,.
}
A similar expression can be found in terms of right-moving variables. 

In contrast to the standard form of the Virasoro-Shapiro amplitude,  
\eqref{VS} is not symmetric in the $s$-, $t$- and $u$-channel. 
Channel duality can be retained by requiring $n_{14}=n_{24} =n_{34}$,
which due to  \eqref{ncond}  implies $n_{ij}=0$. 
In the following, we will argue for this constraint in a more rigorous 
fashion.

\subsubsection*{Pole structure and the strong constraint}

In string theory the poles of the 4-tachyon amplitude appear where 
physical states become on-shell. Thus, they encode the 
mass spectrum of the theory. Now, $\Gamma(x)$ has no zeros but single poles 
at $x=-n$ for $n\in\mathbb{N}$ with residue $\frac{(-1)^n}{n!}$. Therefore the $n^\mathrm{th}$ pole in the $s$-channel is located at
\eq{
	s = \frac{4}{\alpha'}(n-1) \quad\Longleftrightarrow\quad \mathfrak{s}=\frac{4}{\alpha'}(n+n_{34}-1)\,.
}
Hence, we can consider $s=-(k_{L3}+k_{L4})^2 \equiv -(k_L^\mathrm{int})^2$ with $k_{L/R}^\mathrm{int}=p^\mathrm{int}\pm w^\mathrm{int}$ as describing a physical intermediate state with mass and level-matching condition given by
\eq{
	(M^{\mathrm int})^2=-\left((p^\mathrm{int})^2 + (w^\mathrm{int})^2\right) 
	= \frac{4}{\alpha'}\bigl(n+\frac{n_{34}}{2}-1\bigr)
	\!\quad\mathrm{and}\quad\!
	p^\mathrm{int}\!\cdot w^\mathrm{int} = \frac{n_{34}}{\alpha'}\, ,
}
respectively. This corresponds to an asymmetrically excited state with the
difference between the number of right- and left-excitations being $n_{34}$.
However, the condition \eqref{micond} for
modular invariance forbids asymmetrically excited states. Since the same
argument holds for the $t$- and $u$-channel, consistency of the poles
with the physical spectrum requires $n_{ij}=0$. This is nothing else than
the \emph{strong constraint} (in momentum space)
\eq{\label{strong}
	p_i\cdot w_j+p_j\cdot w_i=0 \quad\forall i,j \,.
}
Indeed,  defining the functions as 
$f_i(x,\tilde x)=\exp(ip_i\cdot x+i w_i\cdot \tilde x)$,
the relation \eqref{strong} translates into
\eq{
        \partial_a f_i\, \tilde\partial^a f_j+\tilde\partial^a f_i\,
        \partial_a f_j=0
}
which is the strong constraint \eqref{dSC} of DFT \cite{Siegel:1993th,Hull:2009mi,Hohm:2010jy}.


To summarize, while modular invariance of the partition function 
determined the physical spectrum, consistency with the pole structure of the
Virasoro-Shapiro amplitude allowed to derive the strong constraint.
Let us now combine the  condition \eqref{micond} with the constraint
\eqref{strong}.  In terms of left- and right-moving momenta 
$K_i=(k_{Li},k_{Ri})^t$ the strong constraint reads $\langle K_i,K_j\rangle_d
=0$ $\forall i,j$. Combining it with $k_{R_i}=\mathcal{M}_ik_{Li}$,
 we obtain the joint condition 
\eq{
k_{Li}{}^t\bigl(\mathds{1}-\mathcal{M}_i^t\,\mathcal{M}_j\bigr)k_{Lj}=0
}
which for fixed $i,j$ must hold for all left-moving momenta.
This implies $\mathcal{M}_i=\mathcal{M}_j$ for all $i,j$ so that
both constraints can be summarized by the consistency condition
\eq{\label{extcons}
	k_{Ri} = \mathcal{M}\,k_{Li}\quad\mathrm{with}\,\,\mathcal{M}\in O(d)\,\,\forall i \,.
}
This means that the solution to the strong constraint is chosen
independently of the concrete functions $f,g$ in \eqref{dSC}.


\section{Constraints from torus compactifications}
\label{sec:compact}
In the previous discussion momentum and winding were continuous as we have not
assumed any compact directions. Scherk-Schwarz reductions of DFT 
are examples of configurations   relaxing the strong constraint 
in compact directions \cite{Grana:2012rr}.
The major difference is the quantization of momentum and winding.
It is interesting to see what changes if one repeats our 
analysis from the previous two  sections  
for the case of $k<d$ compact directions.


\subsubsection*{Torus compactification}
We consider general compactifications on a $k$-dimensional torus $T^k=\mathbb{R}^k/2\pi\Lambda_k$ with $\Lambda_k$ a $k$-dimensional lattice. Since the coordinates $X^a$ and $\widetilde{X}^a$ are independent, they can be compactified on different tori $T^k$ and $\widetilde{T}^k$. With indices $I,J,\dots$ indicating the internal directions, the coordinates $X^I$ and $\widetilde{X}^I$ acquire new boundary conditions
\eq{
	X^I(e^{-2\pi i}z,e^{2\pi i}\bar z) &= X^I(z,\bar z) +2\pi\sqrt{\alpha'}\, t^I \\ 
	\widetilde{X}^I(e^{-2\pi i}z,e^{2\pi i}\bar z) 
		&= \widetilde{X}^I(z,\bar z) +2\pi\sqrt{\alpha'}\, \tilde{t}^I
}
with $t^I$ and $\widetilde{t}^I$ vector fields on the internal tori,
i.e. $t\in\Lambda_k$ and $\tilde{t}\in\widetilde{\Lambda}_k$ lattice
vectors. The factors $\sqrt{\alpha'}$ are introduced for
convenience\,\footnote{To make the conventions clear, we point out that for a
  circle they are such that the radius comes with a factor
  $\sqrt{\alpha'}$. Then the internal momentum comes with $\sqrt{\alpha'}$ and
  internal winding with the inverse.}. Using the mode expansion
\eqref{modeexp}, in order to satisfy the boundary conditions, the internal winding and momentum are $w^I = \frac{1}{\sqrt{\alpha'}}\,t^I$ and $p^I = \frac{1}{\sqrt{\alpha'}}\,\tilde{t}^I$ . Then the basic vertex operator \eqref{dtvo} is of the form
\eq{\label{Vc}
	V_{p,w}^{\mathrm{c}}(z,\bar z) &= \noo{e^{ip_\mu X^\mu}\,e^{\frac{i}{\sqrt{\alpha'}}\,\tilde{t}_IX^I}\,
		e^{iw_\mu \widetilde{X}^\mu}\,e^{\frac{i}{\sqrt{\alpha'}}\,t_I\widetilde{X}^I}} \,.
}
Small Greek indices $\mu,\nu,\dots$ now denote the external coordinates. The physical state condition for \eqref{Vc} can be deduced from the conformal weight \eqref{cw:V} as before; it reads $p^\mu w_\mu = -\frac{1}{\alpha'}\,t_I \tilde{t}^I$.
For the $V_{p,w}^{\mathrm{c}}V_{p',w'}^{\mathrm{c}}$-OPE to be single-valued we need $t^I \tilde{t}'_I+\tilde{t}^I t'_I \in\mathbb{Z}$.
Hence the tori are not independent but their lattices are contained in each others dual lattices.

It is convenient to introduce the lattice vectors $t_L^I =\frac{1}{\sqrt{2}}(\tilde{t}^I+t^I)$ and $t_R^I =\frac{1}{\sqrt{2}}(\tilde{t}^I-t^I)$ as well as the bilinear form $\langle\cdot,\cdot\rangle_k$ defined by $\mathrm{diag}(\mathds{1}_k,-\mathds{1}_k)$. With the $2k$-dimensional vector $L=(t_L,t_R)^t$ the above condition for single-valuedness becomes $\langle L,L'\rangle_k \in\mathbb{Z}$.
Denoting the lattice spanned by the $L$'s $\Gamma_{2k}$ this means
$\Gamma_{2k}\subset\Gamma_{2k}^*$, i.e. the lattice is integral. Further restrictions on the lattice $\Gamma_{2k}$ will arise from the partition function.

\subsubsection*{The one-loop partition function}
The partition function can be evaluated as before. The only difference is the zero-mode contribution from the internal momenta and windings. Using \eqref{Zmi} and \eqref{mwint} for the internal part we obtain
\eq{
	Z_\mathrm{c}(\tau,\bar\tau)
	=\frac{V_{d-k}/2}{\bigl(2\pi\sqrt{\alpha'}\bigr)^{d-k}}
\frac{1}{\mathrm{Im}(\tau)^{d-k\over 2}|\eta(\tau)|^{2d}}\,
	\sum_{(t_L,t_R)\in\Gamma_{2k}}e^{i\pi\,t_L^2\,\tau}\,
		e^{-i\pi\,t_R^2\,\bar\tau}\,.
}
Under a modular $T$-transformation, all but the last term is invariant, 
the lattice vectors have to satisfy $\langle L,L\rangle_k \in 2\mathbb{Z}$.
This means that $T$-invariance implies that $\Gamma_{2k}$ has to be an even
lattice. Moreover, using Poisson resummation twice, the partition function is
shown to be invariant under a modular $S$-transformations if
$\Gamma_{2k}=\Gamma_{2k}^*$. 
Hence we have rederived the well known result \cite{Narain:1985jj} that modular invariance requires
the lattice $\Gamma_{2k}$ to be even and self-dual.

Moreover, the external momenta still have to satisfy the condition
\eqref{micond}, i.e.  $k_R^\mu = \mathcal{M}^\mu{}_\nu\, k_L^\nu$ for $\mathcal{M}\in O(d-k)$.
Note that, the physical spectrum in the internal sector is less constrained compared to the non-compact case.

\subsubsection*{Pole structure}
Again we consider the scattering of four vertex operators \eqref{Vc}. The only difference to the analysis in section~\ref{sec:scattering} is that the contractions of momenta and windings split into separate contractions of external and internal momenta and windings. The $n^\mathrm{th}$ pole in the $s$-channel seen from the external point of view is
\eq{
	s^e = \frac{4}{\alpha'}\biggl[n+\frac{1}{2}(t_{L3}+t_{L4})^2-1\biggr]
}
and the difference between the external left- and right-movers is
$\mathfrak{s}^e-s^e=\frac{4}{\alpha'}(n_{34}-\frac{1}{2}\langle
L_3+L_4,L_3+L_4\rangle_k)$. Splitting $n_{34}$ and using level matching allows to write this difference as
\eq{\label{extdiff}
	\mathfrak{s}^e-s^e = \langle K_3^e+K_4^e,K_3^e+K_4^e\rangle_{d-k}
}
with $K_i^e=((k_L^\mu),(k_R^\mu))^t$ collecting the external momenta.
As before, the pole corresponds to an
asymmetrically excited state. However, the external part still has to satisfy the condition \eqref{micond} for modular invariance, i.e. \eqref{extdiff} has to vanish. This implies $\langle K_i^e,K_j^e\rangle_{d-k}=0$, which is equivalent to \eqref{strong}. Then the difference between left- and right-excitations of the intermediate states  is $\langle L_3,L_4\rangle_k$. As asymmetric excitations are valid, this describes a physical state. 
Therefore, the strong constraint still applies to the external directions
whereas no further constraint arises for the internal momenta
and windings.

\section{Conclusion}

In this note we have analyzed a T-duality symmetric CFT, whose tree-level
string scattering amplitudes at the two-derivative level are described by DFT.
From analyzing  one-loop modular invariance
and the pole structure  of the four tachyon amplitude we could deduce that
the strong constraint \eqref{strong} must be imposed 
in all the  {\it non-compact} directions, 
whereas {\it compact}  toroidal directions are not subject to any further constraint
beyond those following from  modular invariance.

These observations are in agreement with the possibility of relaxing the
strong constraint on the internal space in Scherk-Schwarz compactifications
\cite{Grana:2012rr}, in light of which the torus is a special case. The
additional constraints found there 
apply to the possible fluxes, saying that they are constant and  subject
to quadratic constraints. Since fluxes are absent in the torus 
compactifications studied here, no constraints are expected.

It would be interesting to use this CFT approach to study higher order correction to the DFT action.


\bigskip
\bigskip
\noindent
\emph{Acknowledgments:}
We  thank Wan-Zhe Feng for discussion. This work was partially supported by the ERC Advanced Grant "Strings and Gravity"
(Grant.No. 32004) and by the DFG cluster of excellence "Origin and Structure of the Universe".

\appendix
\section{Graviton scattering and the DFT action}
\label{app:gravitonscattering}
In the first part of this appendix we  rederive the on-shell three graviton 
scattering amplitude for vertex operators which do explicitly depend on winding modes in
addition to momenta, given in \cite{Tseytlin:1990va}.
In the second part we are going to expand the DFT action into third order in
fluctuations and show that these interactions precisely match
with the above string scattering amplitude. This computation is meant to
provide evidence for the relevance of this T-duality invariant CFT for DFT. 

\subsection{3-Graviton scattering from CFT}

Calculating an $N$-point function of insertions of graviton vertex operators \eqref{gravVO}
$\mathcal{E}_{p,w}(z,\bar z)$ is combinatorially more involved than
a tachyon amplitude. 
For taking care of that one conveniently defines 
\eq{\label{exponentials}
	\mathcal{V}_i(z_i,\bar z_i) = \,\noo{e^{\kappa_i\cdot\partial X(z_i)
		-\lambda_i\cdot\bar\partial\widetilde{X}(\bar z_i)}
		e^{ip_i\cdot X(z_i,\bar z_i)}\,e^{iw_i\cdot\widetilde{X}(z_i,\bar z_i)}} 
}
with $I$ labeling the winding and momenta and $\kappa_i$, $\lambda_i$
auxiliary parameters. One can  derive the vertex operators corresponding to the
first excited states simply by acting on \eqref{exponentials} with derivatives
with respect to both $\kappa_i$ and $\lambda_i$. This operator is related to a
massless graviton vertex operator $\mathcal{E}_{p_i,w_i}$  by
\eq{
	\mathcal{E}_{p_i,w_i}(z_i,\bar z_i)=E_{iab}\frac{\partial}{\partial\kappa_{ia}}\,
	\frac{\partial}{\partial\lambda_{ib}}\,\mathcal{V}_i\Big\vert_{\kappa_i=\lambda_i=0}\,.
}
The $N$ point correlation function can be written as 
\eq{\label{NVc}
	\corr{\prod_{i=1}^{N}\mathcal{V}_i(z_i,\bar z_i)}=\prod_{1\leq i<j\leq N}
	&|z_i-z_j|^{\alpha'(p_i\cdot p_j+w_i\cdot w_j)}\,
	\Bigl(\frac{z_i-z_j}{\bar z_i-\bar z_j}\Bigr)^{\frac{\alpha'}{2}(p_i\cdot w_j+w_i\cdot p_j)}\,\\
	&\phantom{aaaaaaa}
    \times F_{ij}(z_{ij},\bar z_{ij})\, \delta\Big({\textstyle \sum} p_i\Big)
        \delta\Big({\textstyle \sum} w_i\Big)
}
with
\eq{
	F_{ij}(z_{ij},\bar z_{ij})=\exp\biggl(-\frac{\alpha'}{2}\Bigl[
	\,\,&\frac{\kappa_i\!\cdot\!\kappa_j}{(z_i-z_j)^2}+2i\frac{(p_{[\underline{i}}+w_{[\underline{i}})\!\cdot\!\kappa_{\underline{j}]}}{z_i-z_j}\\
	+&\frac{\lambda_i\!\cdot\!\lambda_j}{(\bar z_i-\bar z_j)^2}
	+2i\frac{(p_{[\underline{i}}-w_{[\underline{i}})\!\cdot\!\lambda_{\underline{j}]}}{\bar z_i-\bar z_j}\Bigr]\biggr)\,.
}
The full 3-graviton amplitude is then given by
\eq{
\mathcal A_{3}(p_i,w_i,&E_i)=g_c^3\,C_{S^2}\, \corr{\prod_{i=1}^{3}(c\, \ov{c}\, \mathcal{E}_{p_i,w_i})}\\
	=&g_c^3\,C_{S^2}\, A(\vec z,\vec{\bar z})\,\prod_{k=1}^{3}E_{kab}\frac{\partial}{\partial\kappa_{ka}}\,
	\frac{\partial}{\partial\lambda_{kb}}\prod_{1\leq i < j \leq 3} F_{ij}(z_{ij},\bar z_{ij})\vert_{\kappa_i=\lambda_i=0}\ ,
}
where $A(\vec z,\vec{\bar z})$ 
collects the contractions of the remaining exponentials \eqref{Ntc}.
Notice that we can treat the derivatives with respect to $\kappa$ and the ones with respect to $\lambda$ separately.
Denoting $F(\vec z,\vec{\bar z}):=\prod_{1\leq i<j\leq 3}F_{ij}(z_{ij},\bar z_{ij})$ and  taking
three derivatives with respect to $\kappa$, we find
\eq{
	\prod_{k=1}^{3}\frac{\partial}{\partial\kappa_{ka}}\,\left.F\right\vert_{\kappa_i=\lambda_i=0}
	&=\frac{\alpha^{'2}}{4}\,\frac{\eta^{ac}k_{1L}^{b}
		+\eta^{bc}k_{3L}^{a}+\eta^{ab}k_{2L}^{c}}{z_{12}z_{13}z_{23}}\\
	&\quad+\frac{\alpha^{'}}{2}\left(\frac{k_{1L}^{a}}{z_{12}}-\frac{k_{3L}^{a}}{z_{23}}\right)
		\left(\frac{k_{2L}^{b}}{z_{12}}+\frac{k_{3L}^{b}}{z_{13}}\right)
		\left(\frac{k_{1L}^{c}}{z_{13}}+\frac{k_{2L}^{c}}{z_{23}}\right)\ ,
}
where we made use of momentum and winding conservation as well as the transverse polarization of $E_{mn}$. The $\lambda$-derivatives can be worked out analogously. 
We can now contract the two parts with the corresponding polarization tensors
of the massless vertex operators to get the full 3-point amplitude. 
We restrict ourselves to second order in momentum and winding and  we consider the correct normalization of the graviton vertex operator which makes it necessary to include a factor of $\frac{2}{\alpha^{'}}$ in each $\mathcal E$. Then we find the 3-graviton scattering amplitude to be
\eq{\label{scatamp}
\mathcal A_{3}(p_i,w_i,E_i)=
	4\pi g_c\,E_{1ad}E_{2be}E_{3cf}t^{abc}\tilde t^{def}+\mathcal{O}(p^{4},p^{3}w,\dots,w^{4})\ ,
}
with
\eq{t^{abc}=\,&\eta^{ca}k_{1L}^{b}+\eta^{ba}k_{2L}^{c}+\eta^{cb}k_{3L}^{a}\\
  \tilde t^{abc}=\,&\eta^{ca}k_{1R}^{b}+\eta^{ba}k_{2R}^{c}+\eta^{cb}k_{3R}^{a}\,.
}
Here we used $C_{S^2}=\frac{8\pi}{\alpha'g_c^2}$ which can be determined from
unitarity  by factorizing the 4-point amplitude \eqref{VS} over the tachyonic pole.
This result was first presented  in \cite{Tseytlin:1990va} and consistently reduces to the well-known 3-graviton scattering amplitude \cite{Scherk:1974ca} for vanishing B-field and zero winding.


\subsection{3-point interaction from DFT}

For our purposes, it is convenient to consider DFT theory formulated in terms of the field $\mathcal E_{ij}=G_{ij}+B_{ij}$ and the dilaton field $d$ \cite{Hohm:2010jy}:
\eq{
S=\int dx\, d\tilde x \, e^{-2d} \Bigl\lbrack-&\frac{1}{4}g^{ik}g^{jl}\mathcal D^{p}\mathcal E_{kl}\mathcal D^{p}\mathcal E_{ij}+\frac{1}{4}g^{kl}(\mathcal D^{j}\mathcal E_{ik}\mathcal D^{i}\mathcal E_{jl}+\bar{\mathcal  D^{j}}\mathcal E_{ki}\bar{ \mathcal D^{i}}\mathcal E_{lj})\Bigr.\\
+&\Bigl.(\mathcal D^{i}d\bar{ \mathcal D^{j}}\mathcal E_{ij} + \bar{ \mathcal D^{i}}d \mathcal D^{j}\mathcal E_{ij}) + 4 \mathcal D^{i}d\mathcal D_{i}d\Bigr\rbrack \ . \label{Eaction}
}	  
Despite the fact that T-duality is no longer a manifest symmetry, this description nicely covers momenta and winding modes in the derivatives $\mathcal D_{i}=\partial_i - \mathcal E_{ik}\tilde \partial^{k}$ and $\bar{\mathcal D_{i}}=\partial_i + \mathcal E_{ki}\tilde \partial^{k}$.
The inverse metric $g^{ij}$ is used to raise indices and we set $2\kappa_d^{2}=1$. 
The construction of this action from string field theory made use of a field
redefinition establishing the link to the low-energy effective field theories
\cite{Michishita:2006dr}.
As given in \cite{Hull:2009mi}, at zeroth order in $\alpha^{'}$ this field redefinition is
\eq{\label{fieldredef}
\mathcal E_{ij}= E_{ij}+ f_{ij}(e,d)\ , \quad f_{ij}(e,d)= e_{ij}+\frac{1}{2}e_{i}{}^{k}e_{kj}+\mathcal O(e^3)\ .
}
Using \eqref{fieldredef},
we now expanding the action \eqref{Eaction} around Minkowski space to cubic
order in the fluctuation $e_{ij}$ (see \cite{Hohm:2010jy}). Here $E_{ij}$ denotes the constant background, which for vanishing $B$-field reduces to the Minkowski metric $\eta_{ij}$. It is important to take the higher order fluctuation into account in the expansion of the different objects. The metric $g_{ij}$ is simply given by $g_{ij}=\frac{1}{2}(\mathcal E_{ij}+\mathcal E_{ji})$ and hence, for example, the expansion of the inverse metric takes the following form
\eq{
g^{ij}=\eta^{ij}-e^{(ij)}+\frac{1}{4}e^{ik}e^{j}{}_{k}+\frac{1}{4}e^{ki}e_{k}{}^{j}+\mathcal O(e^3)\ .
}
Then, up to a total derivative,  the action to cubic order in the fluctuation reads
\eq{
\label{hannover96}
S=\int& dx\, d\tilde x \left\lbrack\frac{1}{4} e_{ij} \Box e^{ij} + \frac{1}{4}(D^{i}e_{ij})^{2} + \frac{1}{4}(\bar D^{j}e_{ij})^{2} -2dD^{i}\bar D^{j} e_{ij} -4d\Box d\right.\\
+&\frac{1}{4} e_{ij}\left((D^{i}e_{kl}(\bar D^{j}e^{kl})-(D^{i}e_{kl})(\bar D^{l}e^{kj})-(D^{k}e^{il})(\bar D^{j}e_{kl})\right)\\
+&\frac{1}{2}d\bigl((D^{i}e_{ij})^{2} +(\bar D^{j}e_{ij})^{2} +\frac{1}{2}(D^{k}e_{ij})^{2} +\frac{1}{2}(\bar D^{k}e_{ij})^{2}\\
+&2e^{ij}(D_{i}D^{k}e_{kj} +\bar D_{j}\bar D^{k}e_{ik})\bigr)
+4\Bigl. e_{ij}dD^{i}\bar D^{j}d+4d^{2}\Box d\Bigr\rbrack\ ,
} 
which was first derived in \cite{Hull:2009mi}. The derivatives are given by 
\eq{
\begin{array}{l}
D_{i}=\partial_i - E_{ik}\tilde \partial^{k}\ , \\
\bar D_{i}=\partial_i + E_{ki}\tilde \partial^{k}\ , 
\end{array}
\quad \Box =&\frac{1}{2} (D^{i}D_{i}+\bar D^{j}\bar D_{j})\ . 
}
In order to compare with the 3-point amplitude  from the CFT side, we 
introduce $\kappa_d$ by  modifying the fluctuation to $2\kappa_d e_{ij}$.
In this way we get a match with the expansion of the standard Einstein-Hilbert
action to third order in the metric fluctuation $h_{ij}$. Then, from the
second line in \eqref{hannover96} and after a partial integration, we identify the interaction term for three $e_{ij}$'s  to be
\eq{
& \kappa_d e_{ij}\Big((D^{i}e_{kl}(\bar D^{j}e^{kl})-(D^{i}e_{kl})(\bar D^{l}e^{kj})-(D^{k}e^{il})(\bar D^{j}e_{kl})\Big)\\
=&-\kappa_d e_{ij}\Big(e^{kl} D^{i}\bar D^{j}e_{kl}+(D^{i}e_{kl})(\bar D^{l}e^{kj})+(D^{k}e^{il})(\bar D^{j}e_{kl})\Big) + (\text{tot. der.})\ .
}
The missing term from the partial integration vanishes because of $D^{i} e_{ij}=0$, following from the polarization constraint as listed in table \ref{tab:states}. 
Next we can read off the value of the 3-graviton vertex in momentum space by 
using $\partial_{i}\rightarrow i p_{i}$ and $\tilde \partial^{i} \rightarrow i
w^{i}$, which translates
derivatives to momenta and winding modes.
Moreover we have to keep track of possible permutations and obtain
\eq{
A^{eee}=4\pi g_c\Bigl(& k_{3R}^{i} e_{1ij}k_{3L}^{j}e_{2}^{kl}e_{3kl}
+k_{3R}^{i}e_{1ij}(e_{2}^{kj})^{T}e_{3kl}k_{2L}^{l}
+k_{3R}^{k} e_{1kl}(e_{3}^{il})^{T}e_{2ij}k_{1L}^{j} \\
&+(\text{cyclic permutations})\Bigr)\ ,
}
where $g_c=\tfrac{\kappa_d}{2\pi}$.
This result nicely matches with the string scattering amplitude
\eqref{scatamp}.The slight difference in the left- and right-moving momenta
can be cured by switching the sign of the $B$-field.


\bibliography{references}  
\bibliographystyle{utphys}

\end{document}